\documentclass[twocolumn]{revtex4-2}
\usepackage[utf8]{inputenc}
\usepackage[T1]{fontenc}
\usepackage{amsmath}

\renewcommand\d{\mathrm{d}}
\newcommand\e{\mathrm{e}}
\DeclareMathOperator\erfc{erfc}
\newcommand\moy[1]{\left\langle{#1}\right\rangle}
\renewcommand\L[1]{\widehat{#1}}

\newcommand\vM{\mathbf{M}}
\newcommand\phit{{p\sb{a}}}
\renewcommand\vr{\mathbf{r}}

\newcommand\vU{\mathbf{U}}
\newcommand\uu{{\hat{\mathbf{u}}}}
\newcommand\vx{\mathbf{x}}
\newcommand\vX{\mathbf{X}}

\usepackage{graphicx}
\begin{document}

\title{Continuous-space event-driven simulations of\\
reaction-diffusion processes in three dimensions}

\author{Vincent Rossetto}
\email[Correspondance: ]{vincent.rossetto@grenoble.cnrs.fr}
\affiliation{Université Grenoble Alpes et CNRS, LPMMC, Grenoble, France}

\begin{abstract}
We show that reaction-diffusion processes in three dimensions
can be efficiently handled by event-driven numerical simulations, based
on statistical waiting times (Gillespie's Monte-Carlo method). 
The algorithm is efficient for dilute systems, since diffusion is
not simulated, only the result of diffusion between events needs to
be implemented.
\end{abstract}
\maketitle

Reaction-diffusion processes are widely studied models in non-equilibrium
physics aiming at describing complex systems in chemistry, biology or
social sciences. Despite their apparent simplicity they reproduce critical
behaviour in systems where a large number of degrees of freedom
are strongly correlated and undergo phase transitions. Since the
critical exponents, among others properties, depend on the number of
space dimensions, dimensionality is a key parameter in reaction-diffusion
processes. When the number of dimensions exceeds a critical value~$d_c$,
\emph{the upper critical dimension},
the mean field approximation becomes exact. 
In dimensions lower than $d_c$ 
the mean field approach fails to capture the correct critical exponents 
of the system, one is lead to perform numerical simulations. 

The diffusive epidemic process (DEP) is a reaction-diffusion process
in which 'healthy' and 'sick' particles can interact. Whenever a sick 
particle interacts with a healthy it may \emph{contaminate} it (turn it into
a sick particle), while sick particles may \emph{heal} (become healthy)
after a characteristic time~$\tau$. The parameter of control is the
ratio $\rho$ of concentration of sick particles. 
DEP has been designed as a model for
population in polluted environment \cite{kree1989}. It was generalized 
for elements with different diffusion coefficients \cite{van_wijland1998}. 
The evolution of~$\rho$ along time is an important feature of this model. 
The mean-field 
analysis states that below a critical value~$\rho_c$, the systems
evolves to an absorbing phase where all particles are 'healthy'. 
The two approaches of DEP mentioned above belong to different universality 
classes regarding the nature of the transition to the absorbing phase
near the critical ratio~$\rho_c$.
The upper critical dimension of DEP is $d_c=4$, so the critical exponents
in physical systems --- one, two or three dimensional --- are
not easily accessible. 

Most of the numerical studies concerning DEP are performed in one dimension
\cite{maia2007,dickman2008} or in two dimensions \cite{bertrand2007} on
lattices with periodic boundary conditions and use a Monte-Carlo algorithm
with updates accounting for the diffusion of particles or the reactions.
Such an approach in three dimensions would require long runs with the
diffusion updates consuming most of the computational power. In this Letter
we propose an algorithm specific to three dimensional 
reaction-diffusion processes where the diffusion is not performed by
random walks on a lattice but only through Brownian statistical laws.

In numerical simulations of complex systems, diffusing particles 
commonly move according to random walks. While this is often necessary,
a better approach is sometimes possible, based on the time statistics
of first occurrence of a relevant event such as, in the case of DEP,
contamination or healing. Using the Brownian statistics, an alternative
algorithm consists in computing the time until the next event occurs,
update the system up to this time, perform the event and repeat. This
procedure is said to be \emph{event-driven} because the systems jumps
from event to event and no updates are dedicated to diffusion only.
It is a extension of the method proposed by Gillespie for numerical simulations
of arbitrary complex master equations~\cite{gillespie1976}.
Event-driven simulations are not only
more efficient but they allow to explore longer time, because in a rarefied
system, when almost no event happen per unit of time, the event-driven
simulations jump to next event, whenever it may happen. 

In this article, we show that in three dimensions, an exceptional situation
allows for event-driven simulations where the diffusing species are spheres.
The algorithm we propose is mostly based on the simple form of the 
statistical laws governing the encounter of two diffusing spheres that 
we compute in the next section. We then describe how the event-driven
algorithm might be implemented.

\section{Time statistics of the meeting\\ of two diffusing spheres}
\begin{figure}[b]
\centering
\includegraphics{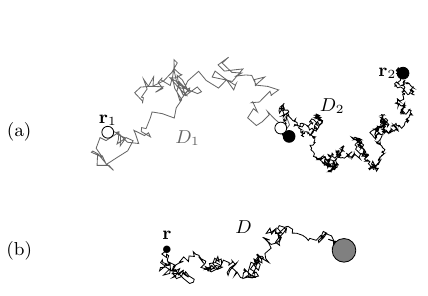}
\caption{\label{2path}
(a) Two Brownian paths starting from $\vr_1$ and $\vr_2$ meeting at point 
$\vx$ with diffusivities $D_1$ and $D_2$ respectively. The position
of the meeting point $\vx$ is distributed as a Gaussian with characteristics
given in the text.
(b) The equivalent Brownian path with diffusivity~$D=D_1+D_2$ starting
from $\vr=\vr_2-\vr_1$. }
\end{figure}
We consider two spheres of radii $a_1$ and $a_2$ diffusing with diffusivities
$D_1$ and $D_2$ respectively. At time~$t=0$, the centers of the spheres,
denoted by $\vr_1$ and $\vr_2$ respectively, 
are separated by a distance $r>a_1+a_2=a$. The position~$\vx$ of a 
particle with diffusivity~$D$ starting from $\vr$ is  
distributed after a time~$t$ according to
\begin{equation}
G_D(\vx,\,t;\,\vr)=\frac{1}{(4\pi Dt)^{3/2}}
  \exp\left(-\frac{(\vx-\vr)^2}{4Dt}\right)
\label{gD}
\end{equation}
To study the meeting problem we consider
the vector $\vX_\vr=\vr_2-\vr_1$, which is a Brownian random 
variable with diffusivity $D=D_1+D_2$ starting with initial value~$\vr$.
The spheres meet whenever $\|\vX_\vr(t)\|=a$.
We denote by $T$ the first occurrence time of this event,
by $a\vU$ the random vector $\vr_2(T)-\vr_1(T)$ and by $\vM$ the
``meeting point'' such that $\vr_1(T)=\vM-a_1\vU$ and $\vr_2(T)=\vM+a_2\vU$.
In the Appendix, we compute the probability that the spheres do not meet between
$t=0$ and the time~$t$ as
\begin{equation}
u(t)=1-\frac{a}{r}\erfc\left(\frac{r-a}{\sqrt{4Dt}}\right). 
\label{u(t)}
\end{equation}
Taking $t$ to infinity,
we retrieve an already known result: The probability that a 3-D Brownian
motion starting at a distance $r$ from a sphere of radius $a$ reaches 
the surface of the sphere \emph{at any time} is $\phit(r)=a/r$. 
We complement this
result by providing that, given that this Brownian eventually reaches the
surface of the sphere, the time of arrival is distributed according
to the probability law
\begin{equation}
P_{T}(t)=\frac{1}{\phit(r)}\frac{\partial u}{\partial t}=
   \frac{r-a}{\sqrt{4\pi Dt^3}}\exp\left[-\frac{(r-a)^2}{4Dt}\right].
\label{pT}
\end{equation}
The expressions of $\phit$ and $P_T$ can be deduced from the
entry \textbf{5}.2.0.2 in Ref.\cite{borodinsalminen}. Note that in two
dimensions, the distribution $P_T$ takes a more complex form
(entry \textbf{4}.2.0.2).
We also obtain the distribution of the first hitting point on the sphere
of radius~$a$ as a Fisher law
\begin{equation}
P_{\vU}({\uu})=\frac{G_D\big(a\uu,\,t;\,\vr\big)}{\gamma_a\big(r,\,t\big)}
=\frac{ar}{8\pi Dt\phantom{\big(}}\frac{\exp\left(\frac{a\vr\cdot\uu}{2Dt}\right)}
                        {\sinh\left(\frac{ar}{2Dt}\right)},
\label{pU}
\end{equation}
where $\gamma_a(r,\,t)$ is the integral 
$\int_{\|\vx\|=a}G_D(\vx,\,t;\,\vr)\d\vx$.

In order to complete our discussion about the two diffusing spheres, 
we add that if they meet (probability $\phit(r)$) and if the
meeting happens after a time $T$ (distributed according to $P_T$, 
Equation~\eqref{pT}) then their meeting position $\vM$ is 
distributed according to the Gaussian distribution 
\begin{align}
P_{\vM}(\vx)&=G_{D'}(\vx,\,T;\,\vr_{12})&\text{with}&\label{pM}\\
\frac{1}{D'}&=\frac{1}{D_1}+\frac{1}{D_2},&
\frac{\vr_{12}}{D'}&=\frac{\vr_1}{D_1}+\frac{\vr_2}{D_2}.
\label{pMcarac}
\end{align}

\section{Event-driven algorithm}
The results from the previous section are the key ingredients to design
an event-driven algorithm for reaction-diffusion processes in three 
dimensions. In this section, we give the characteristics of single particle
reactions and that of two-particle reactions in a general way. Examples
of reaction-diffusion processes are discussed in the next section.

The general principle of an event-driven algorithm is to determinate
from the current configuration the next reaction occurring in the
system and the time when it occurs. The reaction is performed and
the system is made to evolve directly to the time it occurs, updating
the configuration to a new configuration from which the same 
procedure is repeated. This is particularly useful in dilute systems
where most of the computational power is dedicated to simulate the
Brownian movement of all the particles. 

We focus our discussion on one-particle and two-particle reactions.
One-particle reactions are sometimes referred to as ``spontaneous''. 
Extinction and duplication are examples of one-particle reactions:
\begin{align*}
A&\xrightarrow[\quad]{\sigma}2A&\text{(duplication)},\\
A&\xrightarrow[\quad]{\mu}\emptyset&\text{(extinction)}.
\label{reaction1}
\end{align*}
The time until a one-particle reaction occurs is exponentially 
distributed according to (for extinction) 
\[P_E(t)=\mu\exp(-\mu t).\]
Two-particle reactions occur whenever two particles from the same
species or different species approach at a sufficiently close distance
that we will determine. Annihilation and contamination are examples
of two-particle reactions:
\begin{align*}
A+A&\xrightarrow[\quad]{\lambda}\emptyset&\text{(annihilation)},\\
A+B&\xrightarrow[\quad]{k}2B&\text{(contamination)}.
\end{align*}
For every possible pair of particles of radii~$a_1$ and $a_2$, 
we compute the meeting probability
$\phit(r_{ij})=\frac{a_1+a_2}{r}$ where 
$r=\|\vr\|=\|\vr_2-\vr_1\|$ is the distance between their centers.
We first randomly determine if these particles will meet according to
the probability $\phit$. If they meet, we randomly chose the meeting
time~$T$ and the positions of the particles using the distributions
\eqref{pT}, \eqref{pU} and \eqref{pM}. Let us consider for instance the
two-particle annihilation. Since there are $4\pi r^2\rho\d r$ particles at 
distance $r$ from a given particle, the mean-field meeting rate of
two particles is equal to $\int \phit(r) P_T(r) \,4\pi r^2 \rho \d r
\xrightarrow[t\to\infty]{}4\pi D a\rho$
for large times~$t$. This rate is actually equal to the rate $\lambda\rho$,
so we deduce that the radius of interaction~$a$ is determined by
\begin{equation}
a=\frac{\lambda}{4\pi D}. 
\label{a(k)}
\end{equation}
The same reasoning applies if the particles belong to different species.
Any values of $a_1$ and $a_2$ such that $a=a_1+a_2$ determined by \eqref{a(k)}
will result in the correct reaction rate. 
Note that if the same particle species are involved in several reactions,
the reaction constant $\lambda$ in \eqref{a(k)} must be replaced by
$\lambda_{\text{total}}=\lambda_1+\lambda_2+\cdots$. If the
particles meet, the reaction is randomly chosen with probabilities
$\lambda_1/\lambda_{\text{total}}$, $\lambda_2/\lambda_{\text{total}}$, \dots

Interestingly, this event-driven algorithm does not require the introduction
of extra parameters, it is based only on the physical parameters of the system.

\section{Pure annihilation process}
\begin{figure}
\includegraphics[width=0.45\textwidth]{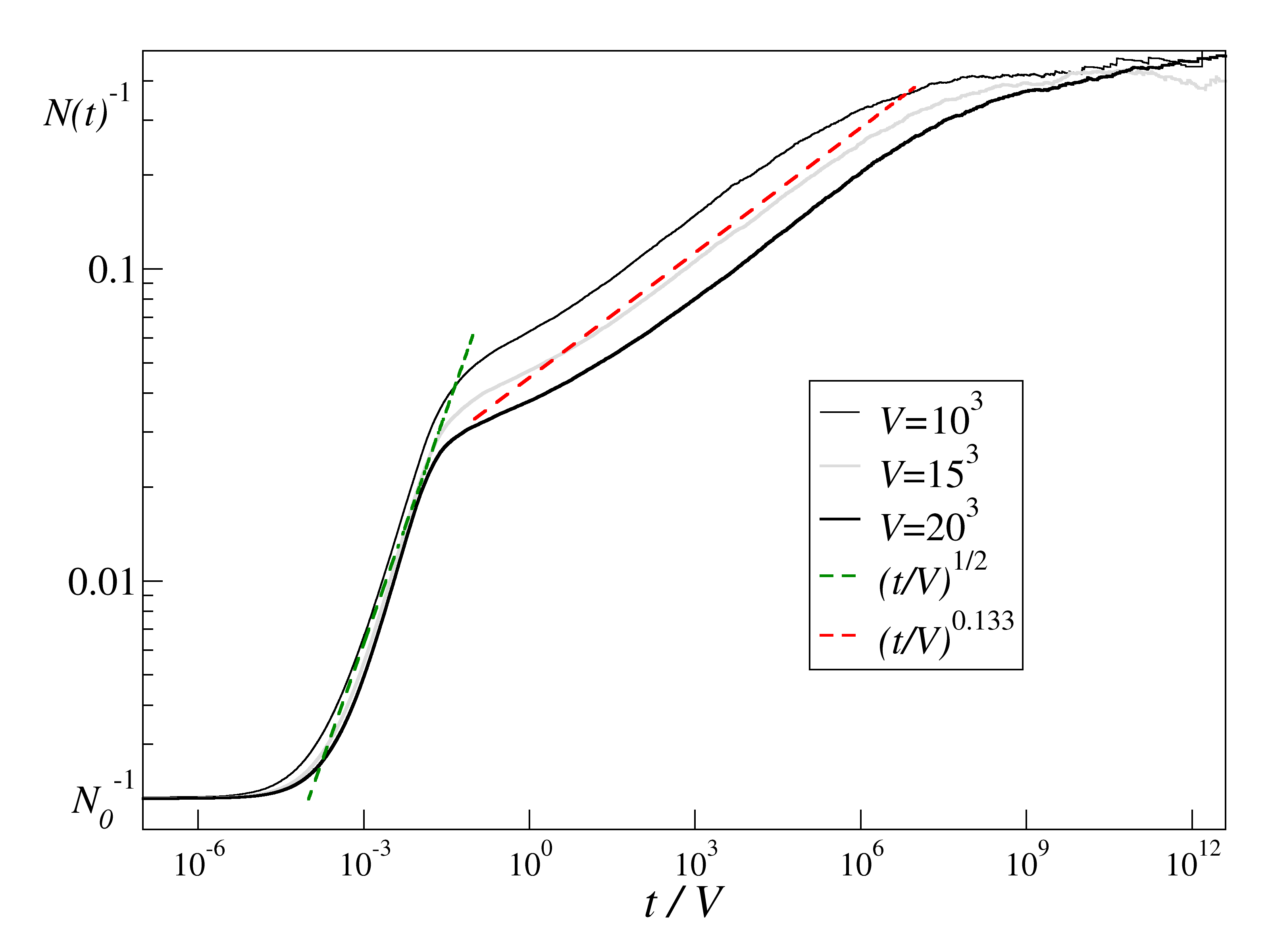}
\caption{\label{fig:annihilation}Average inverse number of particles
as a function of rescaled time for the pure annihilation process with 
$\lambda=2$. }
\end{figure}

We start with the simple process involving only one species and one
annihilation reaction with rate~$\lambda$ in a volume~$V$. 
The mean-field time evolution of the system 
is $\rho(t)^{-1}=\rho_0^{-1}+2\lambda t$. Starting with $N=1000$ particles
in a cubic box of side length~$L=10$, we compute the average number
particles at time~$t$ with $D=1$, $\lambda=2$. The results are displayed
in the Figure~\ref{fig:annihilation}. This example shows that the system
evolves with different dynamics as the mean-field solution. Independently
from the total volume of the reactor, there is a 
rapid diffusive phase where $\rho^{-1} \sim t^{1/2}$, 
and then a
long phase where $\rho^{-1}\sim t^{2/15}$. 

\appendix
\section{Derivation of the probability distributions}
\label{math}
\setcounter{equation}{0}
\renewcommand\theequation{A.\arabic{equation}}
In order compute $P_T$ and $P_{\vU}$ we consider
the \emph{local time} of the random variable $\vX_\vr$. 
The local time density $L(\vX_\vr,\vx,\,t)$ at point~$\vx$ 
is a random variable defined as follows: The time spent by the process
$\vX_\vr$ in the volume $\d^3\vx$ containing $\vx$ during a total 
diffusion time $t$ is distributed like the random variable 
$L(\vX_\vr,\vx,t)\d^3\vx$. It may be expressed as the integral
\begin{equation}
L(\vX_\vr,\vx,\,t)=\int_0^t\delta^{(3)}\big(\vX_\vr(t')-\vx\big)\,\d t'.
\label{Lint}
\end{equation}
The local time on the sphere of radius $a$, denoted by  $L_a(X_r,\,t)$,
depends only on the random variable $X_r=\|\vX_\vr\|$. It is obtained 
by integrating the expression~\eqref{Lint}
with respect to~$\vx$ over the sphere.
Since the expression $\moy{\delta^{(3)}\big(\vX_\vr(t)-\vx\big)}$
is equal to the probability density function of~$\vX_\vr(t)$,
that is to say $G_D(\vx,\,t;\,\vr)$,
the expression of the average local time density is
\[\moy{L_a(X_r,t)}=\int_0^t \iint_{\|\vx\|=a} 
G_D(\vx,\,t';\,\vr)\d t' \d^2\vx.\]

The derivation of the probability density function of~$L_a$ is similar
to the computation of a sensitivity kernel for diffuse waves
\cite{rossetto2013}. Let us denote by $\gamma_a(r,\,t)$ the integral of \eqref{gD} 
with respect to $\vx$ over the sphere of radius~$a$. 
We have $\gamma_a(r,\,t)=\moy{\delta\big(X_r(t)-a\big)}$.
The Laplace transform of $\gamma_a(r,\,t)$ with respect to $t$ is 
\begin{equation}
\L\gamma_a(r,\,s)=\frac{1}{ar}
\frac{\exp\left(-r\sqrt{\frac{s}{D}}\right)}{\sqrt{Ds}}
\sinh\left(a\sqrt{\frac{s}{D}}\right).
\label{gamma}
\end{equation}
Let us now compute the moment of order~$n$ 
$\moy{L_a(X_r,\,t)^n}=\Lambda_n(t)$. It is given by the
$n$-fold integral with integration variables $t_1,\dots,\,t_n$
running from $0$ to $t$ of the integrand
\begin{equation}
\Big\langle
\delta\big(X_r(t_1)-a\big)\;
\delta\big(X_r(t_2)-a\big)\;
\cdots\;
\delta\big(X_r(t_n)-a\big)\Big\rangle.
\label{moy.prod}
\end{equation}
Reordering the times as $0\leq t_1\leq\cdots\leq t_n\leq t$
the integrand gains a prefactor~$n!$ and is
causal: The Brownian path is necessarily on the $a$-sphere at time 
$t_1$ such that from $t_1$ to $t_2$ the random variable $X_r$
starts and ends with the value~$a$. 
Since the Brownian path between $t_i$ and $t_{i+1}$ is independent 
from the Brownian paths during the other time intervals, the factors 
in the product of delta-functions are uncorrelated and $X_r(t_i+\tau)$ 
is under these conditions equivalent to $X_a(\tau)$ for $\tau<t_{i+1}-t_i$.
The integrand~\eqref{moy.prod} rewrites as
\[ n!
\big\langle\delta\big(X_r(\tau_1)-a\big)\big\rangle\;
\big\langle\delta\big(X_a(\tau_2)-a\big)\big\rangle\;
\cdots
\big\langle\delta\big(X_a(\tau_n)-a\big)\big\rangle,
\]
where $\tau_i=t_{i}-t_{i-1}$ (we use $t_0=0$)
with the condition that $\tau_1+\cdots+\tau_n<t$. 
The Laplace transform of $\Lambda_n$ is immediately found as 
\begin{equation*}
\L \Lambda_n(s)=n!\,\frac{1}{s}\L\gamma_a(r,\,s)\L\gamma_a(a,\,s)^{n-1}.
\end{equation*}
We recognize the moments of an non-normalized exponential distribution
$w(s)\exp(-\ell/\L\gamma_a(a,\,s))$, where $w(s)=\frac1s\frac{\L\gamma_a(r,s)}{\L\gamma_a(a,s)}$. 
The normalization --- to $s^{-1}$ in Laplace representation --- is
obtained by adding an extra term of the form~$u(s)\delta(\ell)$ 
with $u(s)=\frac1s-w(s)$~:
\begin{equation}
P_L(\ell)=\left(\frac{1}{s}-\frac{1}{s}
   \frac{\L\gamma_a(r,\,s)}{\L\gamma_a(a,\,s)}\right)\delta(\ell)
 +w(s) \e^{-\ell/\L\gamma_a(a,s)}.
\label{PL}
\end{equation}
In time representation, the regularization coefficient 
$u(t)=1-\frac{a}{r}\erfc\big(\frac{r-a}{\sqrt{4Dt}}\big)$
is the probability that the Brownian motion $\vX_\vr$ does not
hit the $a$-sphere during the total time~$t$, $\erfc$ denoting
the complementary error function. 

\thanks{The author wishes to thank...}

\bibliography{rd3d}
\end{document}